\documentclass[12pt,preprint]{aastex}
\usepackage{lscape}

\newcommand{\bdv}[1]{\mbox{\boldmath$#1$}}

\def\au{{\rm AU}}

\def\mas{{\rm mas}}

\def\rel{{\rm rel}}

\def\geo{{\rm geo}}
\def\e{{\rm E}}
\def\bpi{{\bdv\pi}}
\def\bmu{{\bdv\mu}}

\begin{document}
\title{{\it Spitzer} Observations of OGLE-2015-BLG-1212 Reveal a New
Path
to Breaking Strong Microlens Degeneracies}

\author{
V. Bozza\altaffilmark{D1,D2},
Y. Shvartzvald\altaffilmark{S1,a},
A. Udalski\altaffilmark{O1},
S. Calchi Novati\altaffilmark{D1,D10,S2,b},
I.A.~Bond\altaffilmark{M3},
C. Han\altaffilmark{K1},
M.~Hundertmark\altaffilmark{D3},\\
and\\
R. Poleski\altaffilmark{S3,O1},
M. Pawlak\altaffilmark{O1},
M.\,K. Szyma{\'n}ski\altaffilmark{O1},
J. Skowron\altaffilmark{O1}
P. Mr{\'o}z\altaffilmark{O1},
S. Koz{\l}owski\altaffilmark{O1},
{\L}.~Wyrzykowski\altaffilmark{O1},
P. Pietrukowicz\altaffilmark{O1},
I. Soszy{\'n}ski\altaffilmark{O1},
K. Ulaczyk\altaffilmark{O2},\\
(OGLE group)\\
and\\
C. Beichman\altaffilmark{S2},
G. Bryden\altaffilmark{S1},
S. Carey\altaffilmark{S4},
M. Fausnaugh\altaffilmark{S3},
B. S. Gaudi\altaffilmark{S3},
A. Gould\altaffilmark{S3},
C. B. Henderson\altaffilmark{S1,S3,a},
R. W. Pogge\altaffilmark{S3},
B. Wibking\altaffilmark{S3},
J. C.\ Yee\altaffilmark{S5,c},
W. Zhu\altaffilmark{S3},\\
(Spitzer team)\\
and\\
F.Abe\altaffilmark{M2},
Y. Asakura\altaffilmark{M2},
R.K. Barry\altaffilmark{M13},
D.P. Bennett\altaffilmark{M1},
A. Bhattacharya\altaffilmark{M12},
M. Donachie\altaffilmark{M4},
M. Freeman\altaffilmark{M4},
A. Fukui\altaffilmark{M5},
Y. Hirao\altaffilmark{M6},
K. Inayama\altaffilmark{M7},
Y. Itow\altaffilmark{M2},
N. Koshimoto\altaffilmark{M6},
M.C.A. Li\altaffilmark{M4},
C.H. Ling\altaffilmark{M3},
K. Masuda\altaffilmark{M2},
Y. Matsubara\altaffilmark{M2},
Y. Muraki\altaffilmark{M2},
M. Nagakane\altaffilmark{M6},
T. Nishioka\altaffilmark{M2},
K. Ohnishi\altaffilmark{M8},
H. Oyokawa\altaffilmark{M2},
N. Rattenbury\altaffilmark{M4},
To. Saito\altaffilmark{M9},
A. Sharan\altaffilmark{M4},
D.J. Sullivan\altaffilmark{M10},
T. Sumi\altaffilmark{M6},
D. Suzuki\altaffilmark{M1},
P.,J. Tristram\altaffilmark{M11},
Y. Wakiyama\altaffilmark{M12},
A. Yonehara\altaffilmark{M7},\\
(MOA group)\\
and\\
J.-Y.Choi\altaffilmark{K1},
H. Park\altaffilmark{K1},
Y. K. Jung\altaffilmark{K1},
I.-G. Shin\altaffilmark{K1},
M. D. Albrow\altaffilmark{K2},
B.-G. Park\altaffilmark{K4},
S.-L. Kim\altaffilmark{K3},
C.-U. Lee\altaffilmark{K3},
S.-M. Cha\altaffilmark{K3,K4},
D.-J. Kim\altaffilmark{K3,K4},
Y. Lee\altaffilmark{K3,K4},\\
(KMTNet group)\\
and\\
M. Dominik\altaffilmark{D4,D5},
U.~G. J{\o}rgensen\altaffilmark{D3},
M.~I. Andersen\altaffilmark{D6},
D.~M.~Bramich\altaffilmark{D7},
M.~J. Burgdorf\altaffilmark{D8},
S.~Ciceri\altaffilmark{D9},
G. D'Ago\altaffilmark{D1,D10},
D. F. Evans\altaffilmark{D11},
R.~Figuera~Jaimes\altaffilmark{D4,D12},
S.-H. Gu\altaffilmark{D13},
T. C. Hinse\altaffilmark{K3},
N.~Kains\altaffilmark{D14},
E. Kerins\altaffilmark{D14},
H. Korhonen\altaffilmark{D15,D3},
M.~Kuffmeier\altaffilmark{D3},
L. Mancini\altaffilmark{D9},
A.~Popovas\altaffilmark{D3},
M.~Rabus\altaffilmark{D16},
S. Rahvar\altaffilmark{D17},
R.~T. Rasmussen\altaffilmark{D18},
G. Scarpetta\altaffilmark{D1,D10},
J. Skottfelt\altaffilmark{D22,D3},
C. Snodgrass\altaffilmark{D21},
J.~Southworth\altaffilmark{D11},
J. Surdej\altaffilmark{D19},
E.~Unda-Sanzana\altaffilmark{D20},
C. von Essen\altaffilmark{D18},
Y.-B.~Wang\altaffilmark{D13},
O. Wertz\altaffilmark{D19},\\
(MiNDSTEp) \\
and\\
D. Maoz\altaffilmark{W1},
M. Friedmann\altaffilmark{W1},
S. Kaspi\altaffilmark{W1},\\
(Wise group)\\
}
\altaffiltext{D1}{Dipartimento di Fisica ``E. R. Caianiello'', Universit\`a di Salerno, Via Giovanni Paolo II 132, 84084 Fisciano (SA),\ Italy}
\altaffiltext{D2}{Istituto Nazionale di Fisica Nucleare, Sezione di Napoli, Italy}
\altaffiltext{D3}{Niels Bohr Institute \& Centre for Star and Planet Formation, University of Copenhagen, {\O}ster Voldgade 5, 1350 Copenhagen K, Denmark}
\altaffiltext{D4}{SUPA, School of Physics \& Astronomy, University of St Andrews, North Haugh, St Andrews KY16 9SS, UK}
\altaffiltext{D5}{Royal Society University Research Fellow}
\altaffiltext{D6}{Niels Bohr Institute, University of Copenhagen, Juliane Maries Vej 30, 2100 K{\o}benhavn {\O}, Denmark}
\altaffiltext{D7}{Qatar Environment and Energy Research Institute(QEERI), HBKU, Qatar Foundation, Doha, Qatar}
\altaffiltext{D8}{Meteorologisches Institut, Universit{\"a}t Hamburg, Bundesstra\ss{}e 55, 20146 Hamburg, Germany}
\altaffiltext{D9}{Max Planck Institute for Astronomy, K{\"o}nigstuhl 17, 69117 Heidelberg, Germany}
\altaffiltext{D10}{Istituto Internazionale per gli Alti Studi Scientifici (IIASS), Via G. Pellegrino 19, 84019 Vietri sul Mare (SA), Italy}
\altaffiltext{D11}{Astrophysics Group, Keele University, Staffordshire, ST5 5BG, UK}
\altaffiltext{D12}{European Southern Observatory, Karl-Schwarzschild-Str. 2, 85748 Garching bei M\"unchen, Germany}
\altaffiltext{D13}{Yunnan Observatories, Chinese Academy of Sciences, Kunming 650011, China}
\altaffiltext{D14}{Jodrell Bank Centre for Astrophysics, School of Physics and Astronomy, University of Manchester, Oxford Road, Manchester M13 9PL, UK}
\altaffiltext{D15}{Finnish Centre for Astronomy with ESO (FINCA), V{\"a}is{\"a}l{\"a}ntie 20, FI-21500 Piikki{\"o}, Finland}
\altaffiltext{D16}{Instituto de Astrof\'isica, Facultad de F\'isica, Pontificia Universidad Cat\'olica de Chile, Av. Vicu\~na Mackenna 4860, 7820436 Macul, Santiago, Chile}
\altaffiltext{D17}{Department of Physics, Sharif University of Technology, PO Box 11155-9161 Tehran, Iran}
\altaffiltext{D18}{Stellar Astrophysics Centre, Department of Physics and Astronomy, Aarhus University, Ny Munkegade 120, 8000 Aarhus C, Denmark}
\altaffiltext{D19}{Institut d'Astrophysique et de G\'eophysique, All\'ee du 6 Ao\^ut 17, Sart Tilman, B\^at. B5c, 4000 Li\`ege, Belgium}
\altaffiltext{D20}{Unidad de Astronom{\'{\i}}a, Fac. de Ciencias B{\'a}sicas, Universidad de Antofagasta, Avda. U. de Antofagasta 02800, Antofagasta, Chile}
\altaffiltext{D21}{Planetary and Space Sciences, Department of Physical Sciences, The Open University, Milton Keynes, MK7 6AA, UK}
\altaffiltext{D22}{Centre for Electronic Imaging, Dept. of Physical Sciences, The Open University, Milton Keynes MK7 6AA, UK}

\altaffiltext{O1}{Warsaw University Observatory, Al.~Ujazdowskie~4, 00-478~Warszawa,Poland}
\altaffiltext{O2}{Department of Physics, University of Warwick, Gibbet Hill Road, Coventry, CV4 7AL, UK}

\altaffiltext{S1}{Jet Propulsion Laboratory, California Institute of Technology, 4800 Oak Grove Drive, Pasadena, CA 91109, USA}
\altaffiltext{S2}{NASA Exoplanet Science Institute, MS 100-22, California Institute of Technology, Pasadena, CA 91125, USA}
\altaffiltext{S3}{Department of Astronomy, Ohio State University, 140 W. 18th Ave., Columbus, OH  43210, USA}
\altaffiltext{S4}{{\it Spitzer}, Science Center, MS 220-6, California Institute of Technology,Pasadena, CA, USA}
\altaffiltext{S5}{Harvard-Smithsonian Center for Astrophysics, 60 Garden St., Cambridge, MA 02138, USA}

\altaffiltext{M1}{Laboratory for Exoplanets and Stellar Astrophysics, NASA/Goddard Space Flight Center, Greenbelt, MD 20771, USA}
\altaffiltext{M2}{Solar-Terrestrial Environment Laboratory, Nagoya University, Nagoya 464-8601, Japan}
\altaffiltext{M3}{Institute of Information and Mathematical Sciences, Massey University, Private Bag 102-904, North Shore Mail Centre, Auckland, New Zealand}
\altaffiltext{M4}{Department of Physics, University of Auckland, Private Bag 92019, Auckland, New Zealand}
\altaffiltext{M5}{Okayama Astrophysical Observatory, National Astronomical Observatory of Japan, 3037-5 Honjo, Kamogata, Asakuchi, Okayama 719-0232, Japan}
\altaffiltext{M6}{Department of Earth and Space Science, Graduate School of Science, Osaka University, Toyonaka, Osaka 560-0043, Japan}
\altaffiltext{M7}{Department of Physics, Faculty of Science, Kyoto Sangyo University, 603-8555 Kyoto, Japan}
\altaffiltext{M8}{Nagano National College of Technology, Nagano 381-8550, Japan}
\altaffiltext{M9}{Tokyo Metropolitan College of Aeronautics, Tokyo 116-8523, Japan}
\altaffiltext{M10}{School of Chemical and Physical Sciences, Victoria University, Wellington, New Zealand}
\altaffiltext{M11}{Mt. John University Observatory, P.O. Box 56, Lake Tekapo 8770, New Zealand}
\altaffiltext{M12}{Department of Physics, University of Notre Dame, Notre Dame, IN 46556, USA}
\altaffiltext{M13}{Astrophysics Science Division, NASA/Goddard Space Flight Center, Greenbelt, MD 20771, USA}

\altaffiltext{K1}{Department of Physics, Chungbuk National University, Cheongju 361-763, Republic of Korea}
\altaffiltext{K2}{University of Canterbury, Department of Physics and Astronomy, Private Bag 4800, Christchurch 8020, New Zealand}
\altaffiltext{K3}{Korea Astronomy and Space Science Institute, Daejon 305-348, Republic of Korea}
\altaffiltext{K4}{School of Space Research, Kyung Hee University, Yongin 446-701, Republic of Korea}

\altaffiltext{W1}{School of Physics and Astronomy, Tel-Aviv University, Tel-Aviv 69978, Israel}

\altaffiltext{a}{NASA Postdoctoral Program Fellow}
\altaffiltext{b}{Sagan Visiting Fellow}
\altaffiltext{c}{Sagan Fellow}
\altaffiltext{d}{Royal Society University Research Fellow}

\begin{abstract}

{\it Spitzer} microlensing parallax observations of
OGLE-2015-BLG-1212
decisively breaks a degeneracy between planetary and binary solutions
that is somewhat ambiguous when only ground-based data are
considered. Only eight viable models survive out of an initial set of
32 local minima in the parameter space. These models clearly indicate
that the lens is a stellar binary system possibly located within the
bulge of our Galaxy, ruling out the planetary alternative. We argue
that several types of discrete degeneracies can be broken via such
space-based parallax observations.
\end{abstract}


\section{{Introduction}
\label{sec:intro}}

Strong discrete degeneracies appear generically in the solutions
to microlensing light curves.  Very often these have little practical
importance, either because they are adequately broken by high quality
data or they prove to have very similar scientific implications.
Nevertheless, there are many cases for which an unbroken degeneracy
has serious consequences and thus is quite frustrating, and some
cases (including the one reported here) where it has major
implications for the event in question.  Therefore, any new methods
for breaking these degeneracies deserve the greatest consideration.

To be clear, by ``strong'' degeneracies, we mean those that lead
to very similar light curves over the whole event (or the great
majority of the event).  There is another class of ``accidental''
degeneracies in which the two solutions have very different features
during gaps in the data.  The obvious remedy for the latter is to ensure a full coverage of the microlensing event even in the wings, something that with the advent of new, near-continuous surveys (or combinations of surveys) will become a standard for the great majority of the events. We will therefore concentrate on strong degeneracies, which produce light curves that are indistinguishable from ground-based observatories and thus cannot be solved just by increasing the sampling rate or the coverage.

From theoretical considerations, there is good reason to expect that
observations from a ``microlens parallax satellite'' in solar orbit
might play a powerful role in breaking such degeneracies.  We
illustrate this expectation by considering the most deeply understood
degeneracy: the so-called ``wide/close'' binary degeneracy.  This is
a degeneracy between binary solutions for which the companion lies
outside the Einstein ring ($s>1$, where $s$ is the projected
separation normalized to the angular Einstein radius $\theta_\e$) and
solutions for which the companion is inside the ring $(s<1)$.  This
$(s\leftrightarrow s^{-1})$ degeneracy was discovered empirically in
data for MACHO-98-SMC-1
(Figure~8 from \citealt{ms9801}) at roughly the same time that
\citet{griest98} and \citet{dominik99} derived its fundamental cause:
a deep symmetry between a tidal expansion of the lens equation in the
limit $s\gg 1$ and a quadrupole expansion in the limit $s\ll 1$.
Nevertheless, even though this symmetry is exact in these limits, in
the practical example of MACHO-98-SMC-1 it was already clear that the
full two dimensional (2-D) caustic structure differed significantly
for the two cases. That is, even though the 2-D caustic structures
looked manifestly different, the light curves generated by 1-D tracks
through this structure were virtually identical.  After this same
behavior was noticed for MACHO-99-BLG-47 \citep{mb9947} (see
especially their Figure~4), \citet{an05} was able to explain the
apparent relative ``rotation'' of the two caustics by pursuing the
expansion of the lens equation in each limit to second order (see
also \citealt{Boz00}).

The potential for a parallax satellite to break such degeneracies
{bf can be} recognized by considering Figure~8 of \citet{ms9801} or
Figure~4 of \citet{mb9947}.  This degeneracy arises because the
magnification patterns of the two solutions differ only by an overall
scale factor along the source trajectory, but deviate considerably from
this single scale factor away from this trajectory.  Observing the
event from a satellite introduces a second source trajectory that
probes a different part of the magnification pattern.
For simplicity, consider first that the satellite is not moving with
respect to Earth.  Then the apparent source trajectory through the
caustic structure as seen from the satellite is perfectly parallel to
that seen from Earth but is offset by a 2-D vector that (together
with the known Earth-satellite separation) essentially determines the
parallax vector $\bpi_\e$. Then any physical offset between observatories
will produce caustic crossings in the second trajectory
at distinctly different times for each caustic, exactly because they
are rotated, thereby distinguishing between the solutions. The only
exception would be if the offset were exactly along the trajectory (i.e.,
the source motion is along the Earth-satellite axis), so that there
would be identical lightcurves, just displaced in time.  The same
argument
applies even though the Earth-satellite projected separation changes
with time.  In this more general case, the trajectories are not
perfectly parallel, but they are still rigidly determined (and
separated) for
any fixed choice of $\bpi_\e$.

\begin{figure}
\centering
\includegraphics[width=6cm]{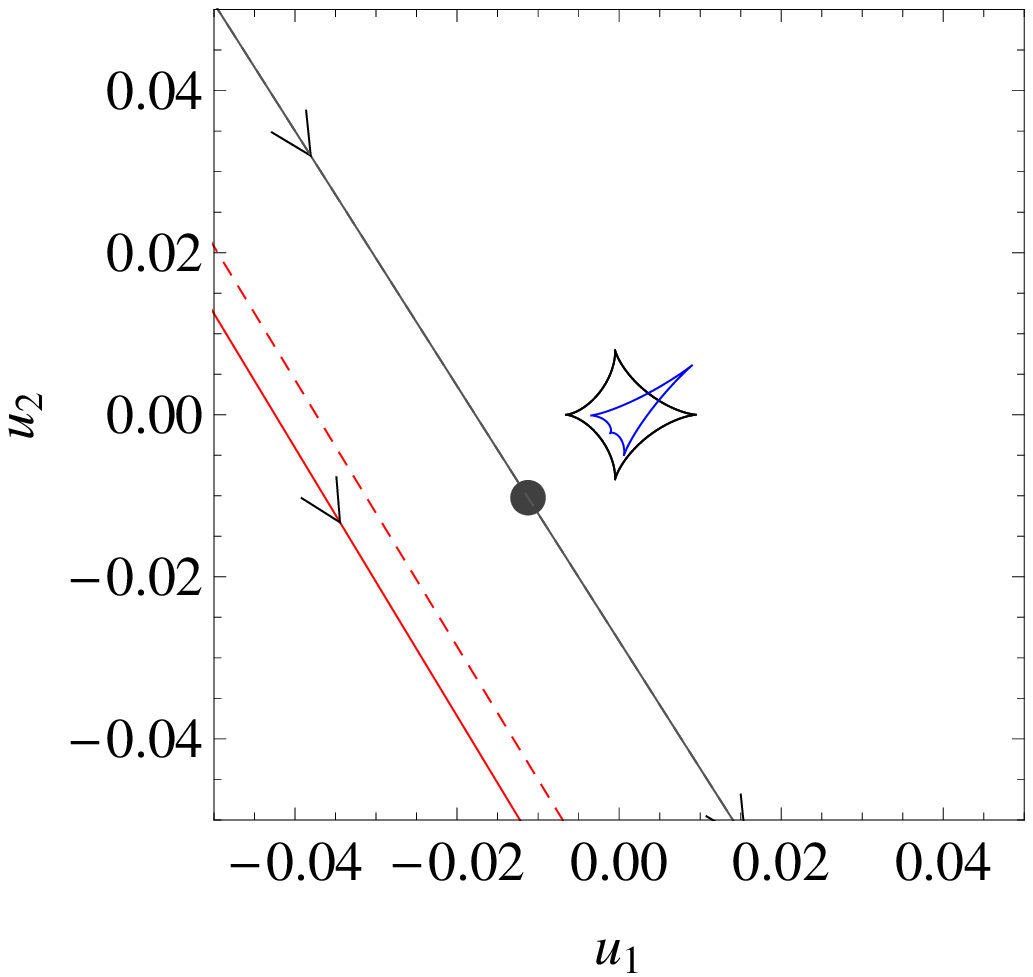}
\caption{Caustics of the best binary model (in black) and the best planetary model (in blue). Also displayed are the source trajectories for the two models: in solid style for the binary and dashed for the planetary; grey for the source as seen from ground observatories, red as seen from Spitzer. In order to compare the two models, the planetary one has been rotated and re-scaled so as to make the source trajectories match as seen from ground  (in practice, they are on top of each other). The corresponding light curves are shown in Fig. \ref{fig:lc} and \ref{fig:compare}. This figure also illustrates the degeneracy discussed by \citet{han08}.}
\label{fig:caust}
\end{figure}

When the source does not experience caustic crossings, the effect of
the binary (or planetary) nature of the lens on the lightcurve is
primarily via cusp approaches.  These can create dramatic bumps
if the source passes close to one or several cusps, but can also
generate weak, longterm distortions for distant passages.
For roughly equal mass binaries, these caustics are
roughly symmetric (concave) quadrilaterals, so that a source can
pass at most two cusps (see Fig. \ref{fig:caust}).  For planets
(with mass ratios $q\ll 1$),
the caustics assume a kite-like form, with three cusps at one end and
one at the other (Fig. \ref{fig:caust}).  On the three-cusp side,
the two outer cusps protrude much further than the central cusp
(which is close to the host).

If, for example, the lightcurve experiences two bumps near its
overall peak, these could in principle be due either to the source
passing two neighboring cusps from a binary caustic or the two outer
cusps from a planetary caustic \citep{han08,choi12,Park14}.  As shown by \citet{han08}, however,
if the passage is close enough, then the shape of the lightcurve
clearly distinguishes between these two cases: for the binary caustic
the interval is rounded, while for planetary caustics, the effect of
the central cusp tends to flatten the intervening lightcurve.
However, for more distant passages, these weaker cusp approaches
open up the possibility of a new
class of degeneracy between central caustics due to binaries and
planets.

Here we present ground and {\it Spitzer} observations of
OGLE-2015-BLG-1212. The ground observations show exactly two such
bumps near peak, which could be due either to a binary or a planet.
In contrast to the cases analyzed by \citet{han08}, however, the
passage is too distant for the central cusp to create obvious
features that would distinguish between the planetary and binary
solutions.  While there remain subtle differences in the models that
permit the ground-based data to distinguish between them at a
moderate level, these are at the level of occasional systematic
effects in microlensing data.  However, the lightcurve obtained by
{\it Spitzer} from its vantage point well displaced from Earth
decisively confirms the preference of ground-based data for the
binary solution.

This result directly impacts the ability of {\it Spitzer}
observations to measure the Galactic distribution of planets, which
is one of the major goals of the {\it Spitzer} microlensing program
\citep{spitzer2015prop}.  As shown by \citet{21event} and
\citet{yee15}, one can determine the relative frequency of planets as
a function of distance from the Galactic center by comparing the
cumulative distribution of planet sensitivity of microlensing events
in the {\it Spitzer} (or other space-based parallax samples) to that
of the planets detected in these surveys.  An implicit assumption of
this approach is, however, that it is known whether a planet is
detected or not, given some specified criteria (e.g.,
$\Delta\chi^2$).  In the present case, the event OGLE-2015-BLG-1212
is high-magnification (and therefore has substantial
sensitivity to planets, \citealt{griest98,gould10}), and has strong
deviations from a \citet{pac86} point-lens lightcurve (meaning that
``something'' has clearly been detected).  However, without breaking
the planet/binary degeneracy, it would not be known whether this
``something'' was a planet.  While it is possible in principle to
take statistical account of such ambiguous cases, they significantly
degrade the statistical power of the experiment, particularly because
the total number of planets detected in space-based microlensing
surveys is small.  Therefore, the fact that {\it Spitzer} itself can
resolve this degeneracy, at least in some cases, adds to its power to
investigate the Galactic distribution of planets.

{\section{Observations}
\label{sec:obs}}

\subsection{OGLE Alert and Observations}

On 2015 June 1, the Optical Gravitational Lensing Experiment \citep[OGLE,][]{OGLE}
alerted the community to a new microlensing event OGLE-2015-BLG-1212
based on observations with the 1.4 deg$^2$ camera on its
1.3m Warsaw Telescope at the Las Campanas Observatory in Chile
using its Early Warning System (EWS) real-time event detection
software \citep{ews1,ews2}.  Most observations were in $I$ band, but
with eight $V$ band observations during the magnified portion of the
event to determine the source color. At equatorial coordinates
(17:52:24.79, $-29$:10:52.0), and Galactic coordinates
$(0.56,-1.40)$, this event lies in OGLE field BLG500, which implies
that it is observed at roughly hourly cadence.

\subsection{{\it Spitzer} Observations}

\citet{ob150966} have reviewed how the {\it Spitzer} team applied
the strategy outlined in \citet{yee15} to select {\it Spitzer}
targets, so we do not repeat those discussions here.  We just summarize
that OGLE-2015-BLG-1212 was ``subjectively'' chosen for observations
on June 7 UT 23:49 (HJD=7181.498), shortly before the Monday upload.  It was
assigned daily cadence but was observed about twice per day that week
(beginning Thursday) due to a general shortage of targets near the
beginning of the program.  The following Monday (June 15) it was
found to meet the objective criteria for a rising event
(\citealt{yee15} criteria ``B''), meaning that all planets discovered
from before the ``subjective'' alert could be incorporated into the
sample, provided that a microlens parallax could be measured from the
post-objective-alert observations. At this time, the $1\,\sigma$
lower limit for the magnification during the next observing interval
was predicted to be $A>80$, which triggered an increase in the
cadence to 8-per-day.  The following week, the
event returned to normal cadence, after which {\it Spitzer}
observations were halted under provisions (``C'') specified by
\citet{yee15}, essentially that the ground-based lightcurve was well
outside the Einstein ring. In fact, this decision was triggered by an
erroneous estimate of the Einstein timescale $t_\e\sim 8\,$days based
on automated point-lens fits to what was in fact a subtly anomalous
lightcurve.  Nevertheless, since the erroneous fit reflected the true
brightness evolution (even though the wrong Einstein-ring position),
it accurately foretold when the target would be too faint to usefully
observe, so that there was no loss of useful observations.
Altogether, {\it Spitzer} observed this event a total of 90 times,
each with 6 dithered 30 s exposures \citep{170event}

\subsection{Other Survey Observations}

\subsubsection{MOA Observations}

Microlensing Observations in Astrophysics (MOA) independently
identified this event on 16 June and monitored it as MOA-2015-BLG-268
using their 1.8m telescope with $2.2\,\rm deg^2$ field at Mt.\ John
New Zealand.  In contrast to most other observatories, which observe
in $I$ band, MOA observes in a broad $R$-$I$ bandpass.  The MOA
cadence for this field is 15 minutes.

\subsubsection{KMTNet Observations}

The event lies in one of four $4\,\rm deg^2$ fields monitored by
Korea Microlensing Telescope Network (KMTNet \citep{KMTNet}) with roughly 15 minute
cadence from its three 1.6m telescopes at CTIO/Chile, SAAO/South
Africa, and SSO/Australia.  Most KMTNet observations are in $I$
band, although some $V$-band observations are taken to determine the
source color. The latter are not used in the present case, as being of poorer quality.

\subsubsection{Wise Observations}

The event lies inside the Wise microlensing survey footprint,
which typically uses the 1m telescope at Wise Observatory, Israel \citep{ShvMao12}.
Due to readout electronics problems with the 1m telescope camera,
as an alternative the Wise group used the Wise C18 0.46m telescope
to monitor the survey fields, including OGLE-2015-BLG-1212.
Observations were in $I$-band, with a cadence of $\sim$1/hour.

\subsection{Followup Observations}

In general, the protocols of \citet{yee15} discourage followup
observations of events with the extremely dense survey coverage
listed above, simply because there are more {\it Spitzer}
events without dense survey coverage than can be adequately covered
by available followup telescopes.  However, the high magnification
(hence, high planet sensitivity) of OGLE-2015-BLG-1212 attracted
dense coverage from several followup groups, particularly over the double peak.

\subsubsection{$\mu$FUN CTIO Observations}

The Microlensing Follow Up Network ($\mu$FUN) observed
OGLE-2015-BLG-1212 using the dual channel ANDICAM camera mounted on
the 1.3m SMARTS telescope at CTIO. Observations started with one point at HJD=7186.9 and ended at 7190.8, concentrating on the last two nights covering the double peak hourly. Most of the optical-channel observations were in $I$ band, with 4 $V$-band observations taken
near peak in order to determine the source color.  All of the
infra-red channel data were in $H$-band. These, again, are primarily
intended for source characterization and are not included in the
fit.

\subsubsection{MiNDSTEp Observations}

The MiNDSTEp consortium observed OGLE-2015-BLG-1212 using the first
routinely operated multi-color instrument mounted on the Danish 1.54
m telescope at La Silla and providing Lucky Imaging photometry. The
instrument itself consists out of two Andor iXon+ 897 EMCCDs and two
dichroic mirrors splitting the signal into a red and a visual part
\citep{skottfelt15}. Observations started at HJD=7189.6 and were continued until 7194.8 with 90 minutes cadence.

\subsection{Data Reduction}

All ground based data were reduced using image subtraction
\citep{alard98} except for the $\mu$FUN CTIO data, which was reduced
with DoPhot \citep{dophot}. The {\it Spitzer} data were reduced with
a new algorithm specifically developed for the {\it Spitzer}
microlensing campaign \citep{170event}. For the analysis of this event, we used the light curve generated by method 3, as explained in that paper.

{\section{Lightcurve Analysis}
\label{sec:anal}}

The basic code used for the calculation of binary microlensing light
curves is the optimized contour integration routine developed by
\citet{Bozza10}. Since there is no caustic crossing, a detailed limb
darkening treatment is unnecessary for this event, and we can proceed
assuming a uniform brightness profile (we have also explicitly checked that the conclusions are unchanged including limb darkening). A preliminary wide search in
the parameter space has been performed by the RTModel software
\footnote{http://www.fisica.unisa.it/gravitationAstrophysics/RTModel.htm},
designed so as to interpret events in real time. After the best
preliminary model has been obtained, we have re-normalized all error
bars so that the total $\chi^2$ equals the number of degrees of
freedom in the fit. More in detail, each dataset has been
re-normalized so that its individual contribution to the $\chi^2$
is proportionate to the number of data points. We remind that the underlying assumption of this procedure is that the noise of all datasets is gaussian in nature.

\begin{figure}
\centering
\includegraphics[width=16cm]{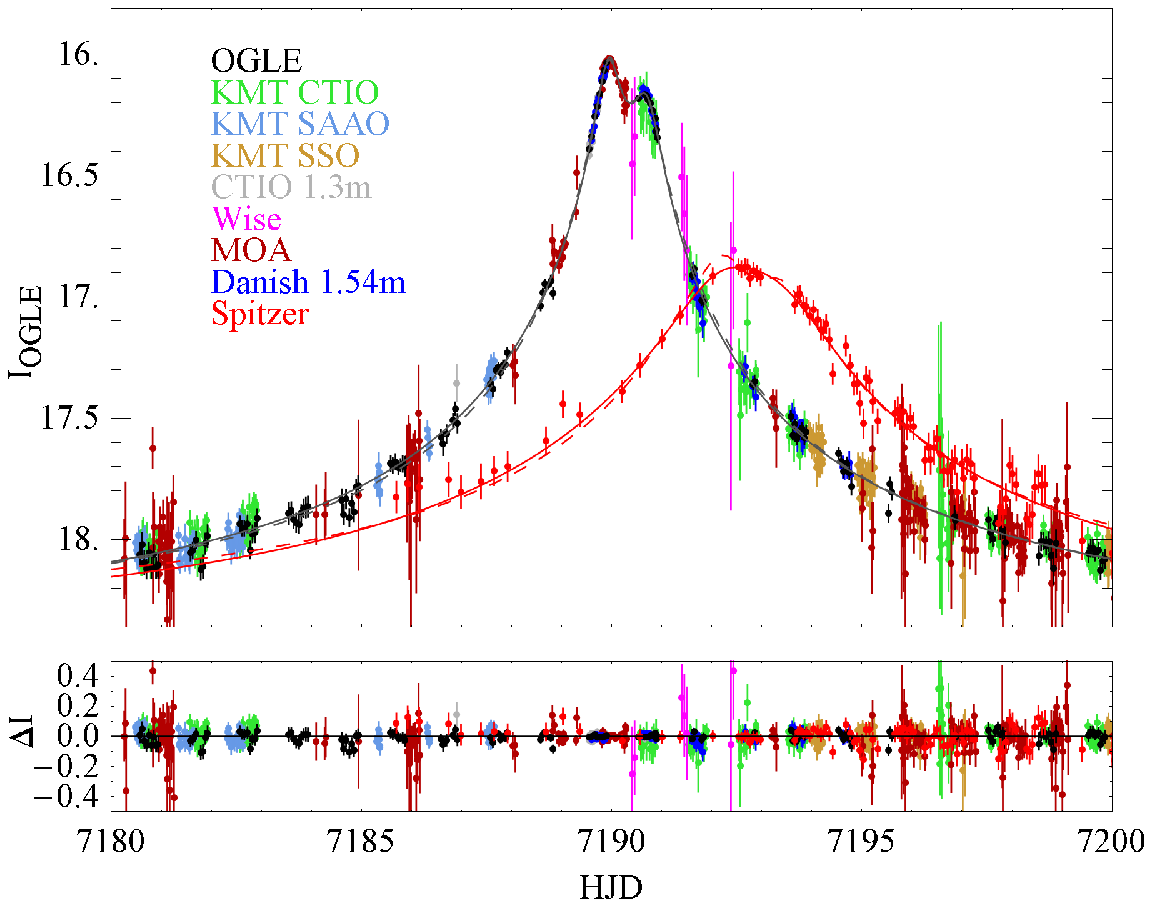}
\caption{Lightcurve of OGLE-2015-BLG-1212 together with the best
binary (solid) and planetary (dashed) models, whose parameters are given in Table 1.
}
\label{fig:lc}
\end{figure}

The light curve of OGLE-2015-BLG-1212 can be obtained by several lens
configurations. In particular, we have identified several solutions
in the planetary regime ($q \lesssim 0.01$) and others in the stellar
binary regime ($q \gtrsim 0.01$). Figure~\ref{fig:lc} shows the light curves
obtained from all the observatories together with the best binary
and planetary models.  The
magnitude scale corresponds to the calibrated $I$-band
magnitudes of the OGLE data. For all other observatories the
magnitudes shown actually represent the magnification, i.e.,
equal ``magnitudes'' at different observatories represent equal
inferred magnifications. Fig. \ref{fig:caust} shows the corresponding caustic structures and
source trajectories
(as seen from Earth and {\it Spitzer}) for the two cases.

Table 1 gives the model parameters for the two solutions shown in
these figures. $u_0$ and $t_0$ are referred to the closest approach to the center of mass of the lens. Note that the planetary solution comes with a mass ratio of 0.002, which, depending on the primary mass, would correspond to a giant planet similar to Saturn. This fact makes this event an extremely interesting study case to test the ability of Spitzer to distinguish between a possible planetary discovery and a simple stellar binary.  As we explain below, each of the these two solutions
is representative of a group of possible solutions, but it is
important to begin by understanding these representative solutions
first.

\begin{table}
\begin{tabular}{ccc}
\hline & Binary & Planetary \\
\hline
$s$ & 0.1760 & 1.5463 \\
$q$ & 0.174 & 0.002423 \\
$u_0$ & -0.01487 & -0.01488 \\
$\theta$ & 2.1386 & 1.4454 \\
$\rho$ & 0.0025 & 0.0019 \\
$t_E$ & 40.22 & 43.6 \\
$t_0$ & 7190.1980 & 7190.2313 \\
$\pi_\perp$ & -0.0639 & -0.0575 \\
$\pi_\parallel$ & -0.0043 & -0.00776 \\
$\chi^2$ & 7952.7 & 8066.1 \\
\hline\end{tabular}
\caption{Comparison of the best binary and the best planetary
solutions. $s$ is the separation between the two lenses in units of the Einstein radius; $q$ is the mass ratio; $u_0$ is the impact parameter to the lens center of mass in Einstein radii; $t_0$ (in HJD) is the time of closest approach to the center of mass, $\theta$ is the angle (in radians) between the source velocity at time $t_0$ and the lens axis, $t_E$ (in days) is the Einstein time; $\rho$ is the source radius in units of the Einstein radius.}
\end{table}

Modeling a caustic-crossing binary (or planet) requires at least
seven geometric parameters to specify the magnification $A(t)$ as a
function of time. The first three $(t_0,u_0,t_\e)$ are the same as
for a single-lens event, namely the time of closest approach by the
source to some fiducial point of the lens geometry (e.g., the center
of mass), the impact parameter (in units of the angular Einstein
radius $\theta_\e$) and the time required to cross the Einstein
radius, i.e.,
\begin{equation}
t_\e \equiv {\theta_\e\over\mu_\geo};
\qquad
\theta_\e\equiv \sqrt{\kappa M\pi_\rel}; \qquad
\kappa \equiv {4G\over\au\,c^2}\simeq 8.1\,{\mas\over M_\odot}.
\label{eqn:tedef}
\end{equation}
Here $M$ is the total lens mass, $\mu_\geo$ is the lens-source
relative proper motion in the geocentric frame and $\pi_\rel
\equiv \au(D_L^{-1}-D_S^{-1})$ is the lens-source relative parallax.  Note
that only the parameter combination $t_\e$ enters the model at this
stage, not the three physical parameters $(M,\mu_\geo,\pi_\rel)$ that
determine it.

The next three parameters $(q,s,\alpha)$ describe the relation of the
primary to the secondary component of the binary.  These are their
mass ratio and their two dimensional separation
$(s\cos\alpha,s\sin\alpha)$
relative to the lens-source trajectory.  Finally, if the source
passes over or near a ``caustic'' (closed curve of infinite
magnification), then the lightcurve profile is smeared out according
to $\rho\equiv\theta_*/\theta_\e$, i.e. the ratio of the angular
source radius to the Einstein radius.

Some events (including all events that, like OGLE-2015-BLG-1212, are
observed from a second observatory in solar orbit) require two
additional parameters, the microlens parallax
\begin{equation}
\bpi_\e = {\pi_\rel\over \theta_\e}{\bmu\over\mu}.
\label{eqn:piedef}
\end{equation}
The numerator of $\pi_\e$ gives the amplitude reflex deflection
of the lens-source apparent position due to displacement by the observer of
1 AU, while the denominator tells the size of
this deflection relative to the Einstein radius, which is what
determines the impact on the light curve. The direction of motion
($\bmu/\mu$) is required to specify the time evolution of this
effect.

From the overall ground-based lightcurve, it is obvious that this
is a high-magnification event ($u_0\ll 1$), so that the double bump
near peak must be due to the effect of two cusps of a central
caustic. As discussed in Section~\ref{sec:intro}, these may be
either consecutive cusps of a ``binary'' ($q\sim {\cal O}(1)$) lens
or opposite prongs of a ``planetary'' ($q\ll 1$) lens.  These
topologies are shown in Figure~\ref{fig:caust}.
Because the {\it Spitzer} lightcurve is broader, the impact parameter
as seen from {\it Spitzer} must be higher, and this is reflected in
the fact that the model shows {\it Spitzer} peaking at lower
magnification. The {\it Spitzer} light curve also peaks later.  These
two offsets (in $u_0$ and $t_0$) determine the parallax, a relation
that can be approximately represented as
\begin{equation}
\bpi_\e = {\au\over D_\perp}(\Delta\tau,\Delta\beta);
\qquad \Delta\tau = {t_{0,\oplus} - t_{0,\rm sat}\over t_\e};
\qquad \Delta\beta = \pm u_{0,\oplus} - \pm u_{0,\rm sat},
\label{eqn:pieframe}
\end{equation}
where the subscripts indicate parameters as measured from Earth
and the satellite and
$D_\perp$ is the Earth-satellite separation projected on the sky.

As is well known, Equation~(\ref{eqn:pieframe}) implies that for
each geometry (as shown in Figure~\ref{fig:caust}), there are three other candidate solutions
\citep{refsdal66,gould94}. As illustrated in Figure~1 of
\citet{gould94}, the four-fold degeneracy
corresponds to (1) the source passing the lens on its right as seen
from both Earth and the satellite
($\Delta\beta_{++} = |u_{0,\oplus}|-|u_{0,\rm sat}|$), (2) both
passing on its left
($\Delta\beta_{--} = -|u_{0,\oplus}|- (-|u_{0,\rm sat}|)$), and (3,4)
passing on opposite sides
$(\Delta\beta_{+-}, \Delta\beta_{-+})$.  Note that the amplitude of
the parallax is the same for (1,2), and also the same for (3,4), but
different between the two pairs.  These identities are exact in the
approximation of Equation~(\ref{eqn:pieframe}) but broken (usually
weakly) by higher order effects \citep{gould95}.  In the case of
caustic-crossing binaries, this degeneracy can be strongly broken
in some cases \citep{graff02,ob151285}, although it may also
persist,
particularly if there is only one caustic crossing observed from
space
\citep{ob141050}.  In the present case, since there are no caustic
crossings,
we do not expect these degeneracies to be strongly broken.  That is,
the situation is qualitatively similar to the point-lens case.

For the planetary model, there is also the close/wide degeneracy,
which is very common for central caustics as first discussed by
\citet{griest98}.  Hence, for the planetary model, there are a total
of $4\times 2=8$ solutions.  For the binary model, the situation
is more complicated.  As in the planetary case, there are both
close and wide models \citep{dominik99,Boz00}.  However, because
the light curve features are in this case due to passage of
consecutive (rather than ``opposite'') cusps of the quadrilateral
caustic, there are in principle four possible orientations for the
caustic for the wide solutions  (compared to one in the planetary
case), and two possible orientations for the close solutions.  See
Figures~2 and 4 of \citet{liebig15}.

In the wide case, these four orientations may be thought of as either
having the companion mass on the same side of the source trajectory
(external cusp approach) or the opposite side (internal cusp
approach), and in each case the companion mass may be passed by the
source either before or after the mass associated with the perturbing
caustic.

In the close case, there are in principle the same four orientations
for the caustic, but the companion mass is always on the same side of
the source trajectory.   Hence, two of these ``different''
orientations actually just represent different mass ratios (i.e.,
$q\rightarrow q^{-1}$), rather than different topologies.  Hence,
there are a total of $4\times 2=8$ close solutions and $4\times 4=16$
wide solutions, and thus $8+8+16=32$ solutions altogether.
These are all shown in Table~\ref{tab:allminima}, with cusp-approach
notation from
\citep{liebig15}.

\begin{table}
\begin{tabular}{cccc}
\hline\hline
 \multicolumn{4}{c}{Close Binary models} \\ \hline  Cusps involved &
 $\Delta\beta_{\pm\pm}$ & $\chi^2$ & $\chi^2$ w/o Spitzer \\ \hline
A--C & $--$ & 7952.7 & 7846.2 \\
A--C & $-+$ & 7955.1 & 7845.8 \\
A--C & $++$ & 7953.0 & 7846.0 \\
A--C & $+-$ & 7953.3 & 7845.7 \\
C--A & $--$ & 8040.5 & 7912.8 \\
C--A & $-+$ & 8040.3 & 7913.3 \\
C--A & $++$ & 8040.3 & 7915.6 \\
C--A & $+-$ & 8040.6 & 7915.9 \\
\hline\hline
 \multicolumn{4}{c}{Wide Binary models} \\ \hline  Cusps involved &
 $\Delta\beta_{\pm\pm}$ & $\chi^2$ & $\chi^2$ w/o Spitzer \\ \hline
A--B & $--$ & 7954.9 & 7858.0 \\
A--B & $-+$ & 7958.6 & 7853.6 \\
A--B & $++$ & 7954.7 & 7851.9 \\
A--B & $+-$ & 7954.1 & 7857.3 \\
B--A & $--$ & 8175.5 & 8042.2 \\
B--A & $-+$ & 8172.0 & 8040.1 \\
B--A & $++$ & 8166.6 & 8053.0 \\
B--A & $+-$ & 8168.5 & 8046.4 \\
D--B & $--$ & 8087.5 & 7984.5 \\
D--B & $-+$ & 8099.8 & 7991.0 \\
D--B & $++$ & 8088.2 & 7986.1 \\
D--B & $+-$ & 8105.2 & 7991.1 \\
B--D & $--$ & 8216.9 & 8100.1 \\
B--D & $-+$ & 8225.8 & 8087.5 \\
B--D & $++$ & 8211.7 & 8089.3 \\
B--D & $+-$ & 8223.5 & 8096.8 \\
\hline\hline
 \multicolumn{4}{c}{Planetary models} \\ \hline  Topology &
 $\Delta\beta_{\pm\pm}$ & $\chi^2$ & $\chi^2$ w/o Spitzer \\ \hline
Close & $--$ & 8066.8 & 7926.0 \\
Close & $-+$ & 8107.1 & 7926.1 \\
Close & $++$ & 8066.8 & 7926.3 \\
Close & $+-$ & 8108.4 & 7925.5 \\
Wide & $--$ & 8066.1 & 7901.5 \\
Wide & $-+$ & 8080.0 & 7898.1 \\
Wide & $++$ & 8066.3 & 7901.3 \\
Wide & $+-$ & 8083.9 & 7898.0 \\
\hline\hline\end{tabular}
\caption{All relevant minima found with a comparison of the $\chi^2$
obtained including or excluding Spitzer data. The notation employed
to indicate the cusps involved in the binary solutions is taken from
Liebig et al. (2015).} \label{tab:allminima}
\end{table}

The values of the $\chi^2$ reported in Table \ref{tab:allminima}
are the final results of extensive Markov chains starting from the
preliminary minima found by RTModel after all possible reflections
discussed above have been applied.

As can be seen from this table, even without {\it Spitzer} data, the
binary solutions are favored over the planetary solutions by
$\Delta\chi^2=53$. This would be regarded as significant evidence for
the binary solution, but not completely compelling due to the
possibility of correlated errors in microlensing data.  See for
example the detailed investigation of one high-magnification event by
\citet{mb10311}, which they argued, are particularly prone to such
systematic errors.  However, {\it Spitzer} data provide independent
evidence of the correctness of the binary solution, raising the total
difference to $\Delta\chi^2=114$.  This seals the case.

\begin{figure}
\centering
\includegraphics[width=16cm]{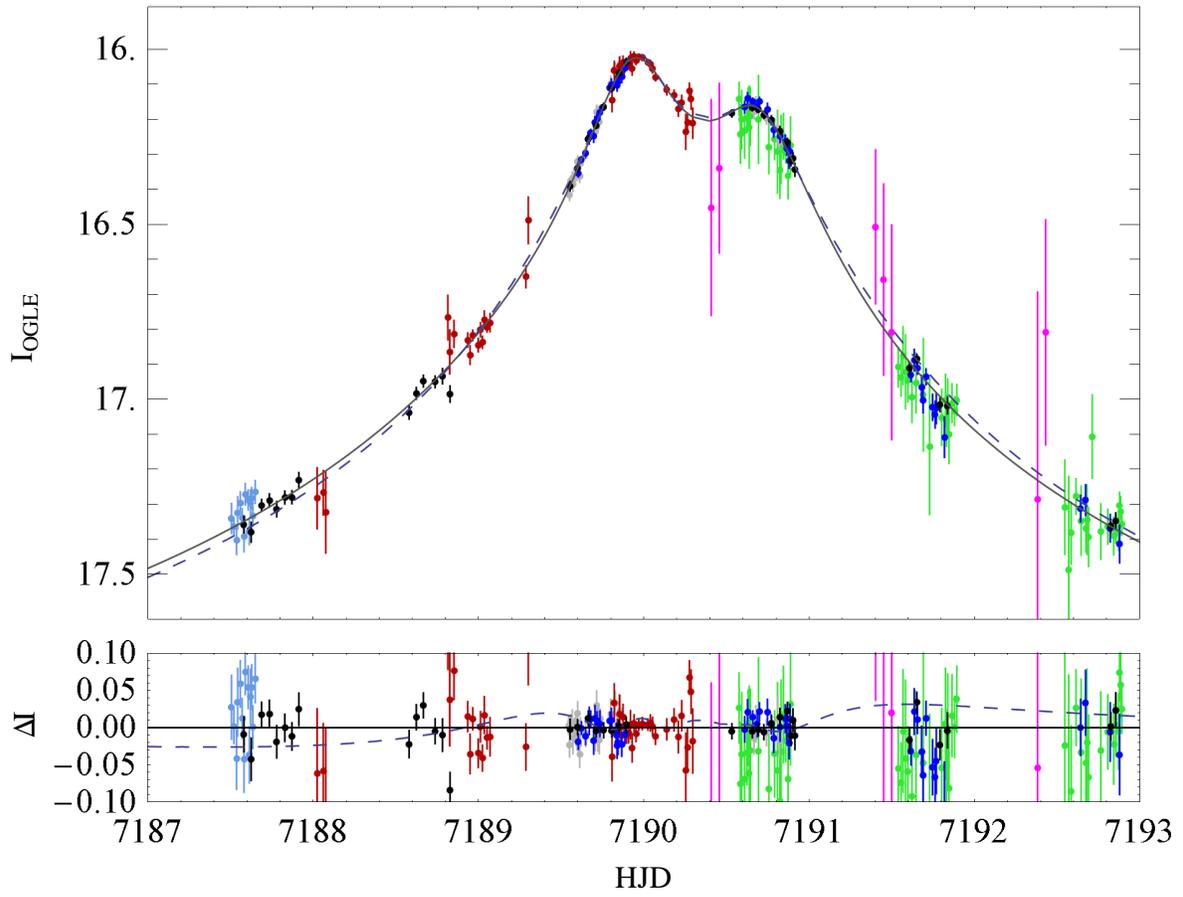}
\caption{Comparison between the binary model (solid curve) and the planetary model (dashed curve) for the data acquired by ground observatories.
}
\label{fig:compare}
\end{figure}

\begin{figure}
\centering
\includegraphics[width=10cm]{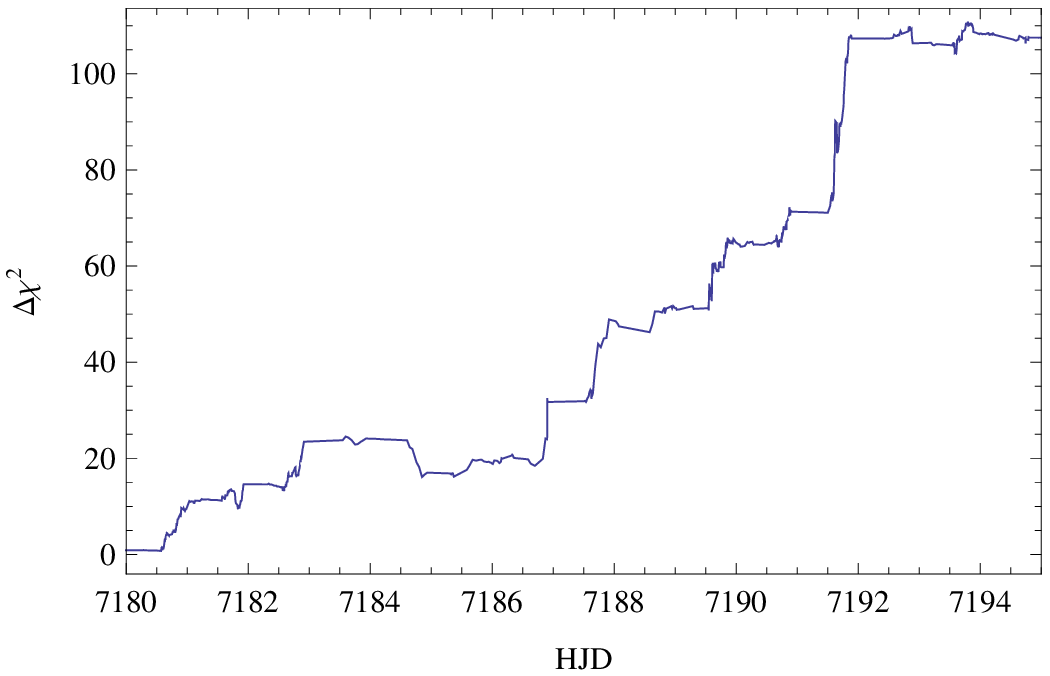}
\caption{$\Delta\chi^2$ between the planetary and the binary model as a function of time.
}
\label{fig:deltachi}
\end{figure}

In Fig. \ref{fig:lc} the planetary model for the ground light-curve is practically indistinguishable from the binary one. However, we can clearly see that the Spitzer lightcurve is different in the two models. While the binary model predicts a smooth slightly asymmetric peak, the planetary model still preserves a concave structure between a main peak and a shoulder. The data point at HJD=7192, however, contradicts the existence of a main peak as suggested by the planetary model. This model also predicts lower magnification during the rising part, being further disfavored.

Fig. \ref{fig:compare} zooms in the peak region as seen from ground observatories comparing the best binary and planetary models. In the double peak region, both models perform quite well. However, we note that before the peak, during the
night $7187.5<HJD<7188$, the data from OGLE,
and KMTNet SAAO are too high above the planetary model, while
after the peak, during the night $7191.5<HJD<7192$, the points from OGLE, Danish and KMTNet CTIO are too low. The binary model fits the data much better.
This discrepancy is the primary origin of
the $\Delta\chi^2=53$ using ground data only. This is also evident from the plot of $\Delta\chi^2$ between the planetary and binary model (Fig. \ref{fig:deltachi}), which shows big steps corresponding to these two nights.
We deduce that the planetary model forces the light curve to have an
asymmetry not reproduced by the data. Nevertheless, the deviations
from the model are still of the order of one sigma and could still be the outcome of some unknown systematics. The contribution by the Spitzer observations is decisive to
discriminate between the two solutions. This
example clearly shows how observations from a different vantage point of the same event are extremely important to correctly classify an
ambiguous microlensing event.

We finally note that the best model fits the data so well that if any systematics are present they are well below the statistical error, thus supporting the work hypothesis of uncertainties dominated by random gaussian noise.

{\section{Physical Character of System}
\label{sec:phys}}

The principal goal of our investigation is to determine whether
the system is planetary or binary in nature because if the
ambiguity remained, this would degrade the measurement of the
Galactic distribution of planets.  That is, the event is very
sensitive to planets, so it is important to determine whether
or not one was detected. As discussed in the previous section, this ambiguity is resolved by the Spitzer data in favor of the binary interpretation.

However, the remaining degeneracies within the binary solution are
quite severe and limit the complete characterization of the system.
Tables 3 and 4 contain full details of the four best close binary
solutions and the four best wide binary solutions respectively. The
close binary solutions arise from the source approaching cusp $A$
(along the lens axis) and then cusp $C$ (off-axis cusp) of the
central caustic. The wide solutions arise from the approach to cusp
$A$ (on-axis) and then $B$ (off-axis) of the perturbed caustic of the
heavier component. These caustics are very similar
\citep{dominik99,Boz00} and generate practically undistinguishable
light curves. In principle, continuing Spitzer observations for some
time after the main event would have probably helped constraining the
existence of a second bump at the closest approach with the caustic
of the secondary object in the wide configuration.

Apart from the wide/close degeneracy, we also have the four-fold
parallax degeneracy discussed in Section \ref{sec:anal}. The symbols
$--$, $-+$, $++$, $+-$ indicate the signs of $u_0$ for the source as
seen from Earth and Spitzer respectively. All these eight solutions
yield a nearly equal $\chi^2$, as can be read from the last lines of
Tables \ref{tab:close} and \ref{tab:wide}, with a very slight
preference for the close models by $\Delta\chi^2 \sim 1$.
Interestingly, all models provide an upper limit for the source
radius parameter $\rho=\theta_*/\theta_E$ of the order of 0.003,
while only those solutions in which the source passes the caustic
from the same side as seen from the Earth and Spitzer (indicated by
the symbols $++$ and $--$) are able to provide a lower limit as well. The
resulting uncertainty is of the order $50\%$, which, combined with
the $4\%$ accurate parallax measurement obtained with Spitzer, is
enough to constrain the lens mass and distance significantly. Note
that both components of the parallax vector are accurately measured,
something that is seldom possible from Earth.

\begin{table}
\begin{tabular}{c|cccc}
\hline
 & \multicolumn{4}{c}{Close} \\ \hline
Parameter & $--$ & $-+$ & $++$ & $+-$ \\
  \hline$s$ & $0.1760_{-0.0062}^{+0.0098}$ &
  $0.1698_{-0.0085}^{+0.0072}$ & $0.1735_{-0.0043}^{+0.0083}$ &
  $0.1690_{-0.0064}^{+0.0097}$ \\
$q$ & $0.174_{-0.023}^{+0.015}$ & $0.188_{-0.018}^{+0.029}$ &
$0.181_{-0.020}^{+0.01}$ & $0.186_{-0.017}^{+0.025}$ \\
$u_0$ & $-0.01487_{-0.00085}^{+0.00031}$ &
$-0.01490_{-0.00064}^{+0.00038}$ & $0.01499_{-0.00047}^{+0.00054}$ &
$0.01466_{-0.0001}^{+0.00092}$ \\
$\theta$ & $2.1386_{-0.0144}^{+0.0091}$ & $2.145_{-0.01}^{+0.013}$ &
$10.4267_{-0.0079}^{+0.0116}$ & $10.422_{-0.012}^{+0.012}$ \\
$\rho_*$ & $0.0025_{-0.0011}^{+0.0019}$ & $<0.0031$ &
$0.001413_{-0.00025}^{+0.00225}$ & $<0.0030$ \\
$t_E$ & $40.22_{-2.31}^{+0.46}$ & $39.99_{-1.74}^{+0.85}$ &
$39.8_{-1.3}^{+1.2}$ & $40.52_{-2.28}^{+0.16}$ \\
$t_0$ & $7190.1980_{-0.006}^{+0.003}$ & $7190.2017_{-0.005}^{+0.005}$
& $7190.1992_{-0.005}^{+0.003}$ & $7190.2005_{-0.004}^{+0.005}$ \\
$\pi_\perp$ & $-0.0639_{-0.004}^{+0.001}$ &
$-0.0347_{-0.0023}^{+0.0014}$ & $-0.0462_{-0.0017}^{+0.0019}$ &
$-0.07640_{-0.0043}^{+0.0008}$ \\
$\pi_\parallel$ & $-0.0043_{-0.001}^{+0.0016}$ &
$-0.0719_{-0.004}^{+0.002}$ & $-0.0445_{-0.0013}^{+0.0022}$ &
$0.0235_{-0.0013}^{+0.0027}$ \\
$\pi_E$ & $0.0640_{-0.001}^{+0.004}$ & $0.0798_{-0.002}^{+0.004}$ &
$0.0641_{-0.002}^{+0.002}$ & $0.07991_{-0.0007}^{+0.0047}$ \\
$\theta_E$ (mas) & $0.22_{-0.12}^{+0.14}$ & $<6$ &
$0.374_{-0.255}^{+0.006}$ & $<5$ \\
$\mu$ (mas/yr) & $2.0_{-1.}^{+1.4}$ & $<61$ & $3.42_{-2.32}^{+0.07}$
& $<46$ \\
$M_1/M_\odot$ & $0.36_{-0.20}^{+0.25}$ & $<8$ &
$0.622_{-0.426}^{+0.003}$ & $<6$ \\
$M_2/M_\odot$ & $0.064_{-0.037}^{+0.042}$ & $<1.7$ &
$0.1081_{-0.0724}^{+0.0027}$ & $<1.2$ \\
$D_L$ (kpc) & $7.18_{-0.6}^{+0.4}$ & $<7.37$ &
$6.6748_{-0.004}^{+0.854}$ & $<7.30$ \\
s (AU) & $0.28_{-0.13}^{+0.14}$ & $<2.07$ & $0.448_{-0.283}^{+0.017}$
& $<2.057$ \\
$I_s$ & $22.081_{-0.067}^{+0.017}$ & $22.068_{-0.044}^{+0.031}$ &
$22.066_{-0.034}^{+0.030}$ & $22.072_{-0.051}^{+0.022}$ \\
$\chi^2$ & $7952.7$ & $7955.1$ & $7953.0$ & $7953.3$ \\
\hline\end{tabular} \caption{The four best close binary solutions
found, with all fit parameters, derived physical parameters and
confidence intervals at $68\%$.} \label{tab:close}
\end{table}

\begin{table}
\begin{tabular}{c|cccc}
\hline
 & \multicolumn{4}{c}{Wide}  \\ \hline
Parameter & $--$ & $-+$ & $++$ & $+-$ \\
  \hline$s$ & $6.59_{-0.30}^{+0.04}$ & $6.6601_{-0.251}^{+0.01}$ &
  $6.839_{-0.33}^{+0.05}$ & $7.089_{-0.32}^{+0.02}$ \\
$q$ & $0.2171_{-0.0294}^{+0.0007}$ & $0.2130_{-0.0234}^{+0.0026}$ &
$0.233_{-0.031}^{+0.001}$ & $0.2455_{-0.0299}^{+0.0017}$ \\
$u_0$ & $-0.97_{-0.02}^{+0.13}$ & $-0.9793_{-0.009}^{+0.117}$ &
$1.063_{-0.14}^{+0.012}$ & $1.1607_{-0.166}^{+0.0086}$ \\
$\theta$ & $2.138_{-0.011}^{+0.002}$ & $2.1384_{-0.0073}^{+0.0051}$ &
$10.4262_{-0.0011}^{+0.0102}$ & $10.41544_{-0.00052}^{+0.01166}$ \\
$\rho_*$ & $0.0032_{-0.0008}^{+0.0013}$ & $<0.0033$ &
$0.0027_{-0.0013}^{+0.0009}$ & $<0.004$ \\
$t_E$ & $44.87_{-2.32}^{+0.04}$ & $44.13_{-1.62}^{+0.55}$ &
$44.3_{-1.6}^{+1.0}$ & $45.08_{-1.91}^{+0.60}$ \\
$t_0$ & $7216._{-4.}^{+1.}$ & $7217.29_{-3.3}^{+0.2}$ &
$7220.58_{-4.7}^{+0.2}$ & $7223.895_{-5.262}^{+0.005}$ \\
$\pi_\perp$ & $0.013909_{-0.00451}^{+0.000062}$ &
$-0.0557_{-0.003}^{+0.002}$ & $-0.0249_{-0.0037}^{+0.0013}$ &
$0.0445_{-0.0031}^{+0.0017}$ \\
$\pi_\parallel$ & $0.0563_{-0.0002}^{+0.0036}$ &
$0.0469_{-0.0026}^{+0.0014}$ & $0.0522_{-0.002}^{+0.001}$ &
$0.05645_{-0.0002}^{+0.0037}$ \\
$\pi_E$ & $0.05765_{-0.0005}^{+0.0033}$ & $0.0728_{-0.002}^{+0.003}$
& $0.0578_{-0.001}^{+0.002}$ & $0.0719_{-0.001}^{+0.003}$ \\
$\theta_E$ (mas) & $<1.031$ & $<4.64$ & $0.206_{-0.073}^{+0.140}$ &
$<8.51$ \\
$\mu$ (mas/yr) & $1.40_{-0.60}^{+4.34}$ & $<39.0$ &
$1.70_{-0.60}^{+1.63}$ & $<75.1$ \\
$M_1/M_\odot$ & $0.30_{-0.13}^{+0.87}$ & $<6.19$ &
$0.35_{-0.11}^{+0.26}$ & $<12.13$ \\
$M_2/M_\odot$ & $0.066_{-0.03}^{+0.038}$ & $<1.27$ &
$0.083_{-0.03}^{+0.053}$ & $<2.776$ \\
$D_L$ (kpc) & $7.41_{-0.3}^{+0.2}$ & $<7.4$ & $7.31_{-0.6}^{+0.2}$ &
$<7.28$ \\
s (AU) & $8.5_{-3.}^{+3.4}$ & $<87$ & $10.3_{-3.6}^{+6.3}$ & $<94$
\\
$I_s$ & $22.0897_{-0.0633}^{+0.0038}$ & $22.056_{-0.035}^{+0.040}$ &
$22.060_{-0.037}^{+0.040}$ & $22.067_{-0.036}^{+0.037}$ \\
$\chi^2$ & $7954.9$ & $7958.6$ & $7954.7$ & $7954.1$ \\
\hline\end{tabular}
\caption{The four best wide binary solutions
found, with all fit parameters, derived physical parameters and
confidence intervals at $68\%$.} \label{tab:wide}
\end{table}

\begin{figure}
\centering
\includegraphics[width=10cm]{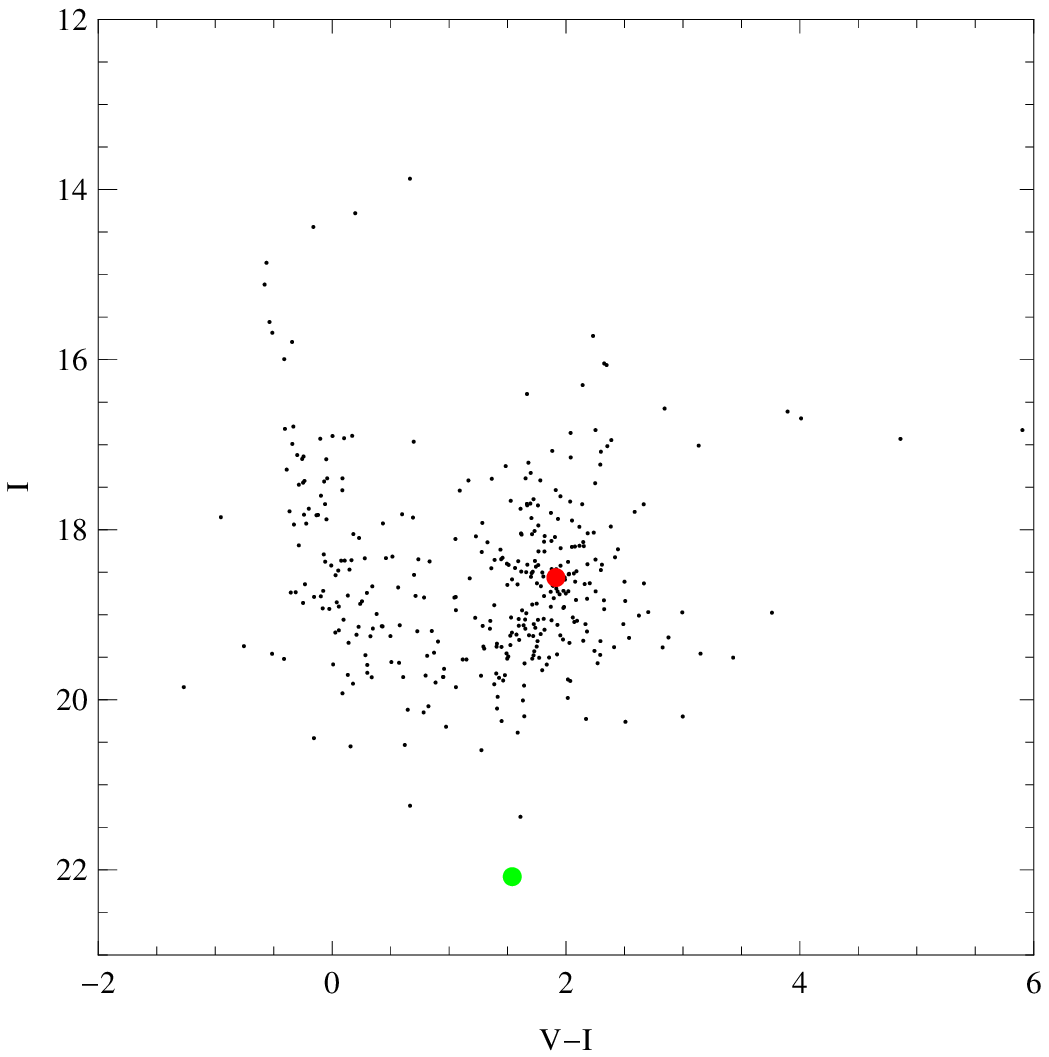}
\caption{Color-magnitude diagram for the field of OGLE-2015-BLG-1212.
The red dot is the centroid of the red giant clump and the green dot
is the position of the source.
}
\label{fig:cmd}
\end{figure}

In order to obtain the physical parameters of the system from the
basic microlensing parameters, we need a complete characterization of
the source involved in the microlensing event. A calibrated $(V,I)$
color-magnitude diagram (CMD) has been obtained by CTIO observations, as
shown in Fig. \ref{fig:cmd}. In particular, the source magnitude is
one of the parameters of the fit, reported in Tables \ref{tab:close}
and \ref{tab:wide}. The source color is obtained by linear regression
on CTIO observations, which have been taken in both colors on the
night $HJD=7189$. We have $V-I=1.54$. After locating the red clump
centroid in the CMD of Fig. \ref{fig:cmd} at $(V-I,I)_{clump} =
(1.93, 18.62)$ and calibrating with $(V-I,I)_{clump,0} = (1.06,
14.42)$ from \citet{nataf13}, we obtain a de-reddened source
$(V-I,I)_{source,0} = (0.67, 17.87)$. This color index translates to
$V-K=1.435$, using the relations in \citet{bb88}. Finally, from
\citet{kervella04}, we find an angular radius $\theta_*\simeq 0.81
\mu$as for the best model. For each point in the Markov chains we can update this value with the parameters of each calculated model and
derive accurate distributions for all secondary physical parameters.

Once we have the angular source radius, we can derive the Einstein
angle, the proper motion, the total mass and the distance from the
formulae
\begin{equation}
\theta_E=\frac{\theta_*}{\rho}; \;\; \mu=\frac{\theta_E}{t_E}; \;\;
M= \frac{AU c^2}{4G} \frac{\theta_E}{\pi_E}; \; \; D_L=
\left(\frac{\theta_E\pi_E}{AU} + \frac{1}{D_S} \right)^{-1}.
\end{equation}

In the table \ref{tab:close} and \ref{tab:wide} we present the
results for these physical parameters. Of course, for those models
for which only an upper limit on $\rho$ is obtained from the light
curve, these parameters are poorly constrained. On the other hand,
for the models for which the source size is well-constrained, we have
relatively small ranges for the masses of the components of the
binary system and for the lens distance.

\begin{figure}
\centering
\includegraphics[width=10cm]{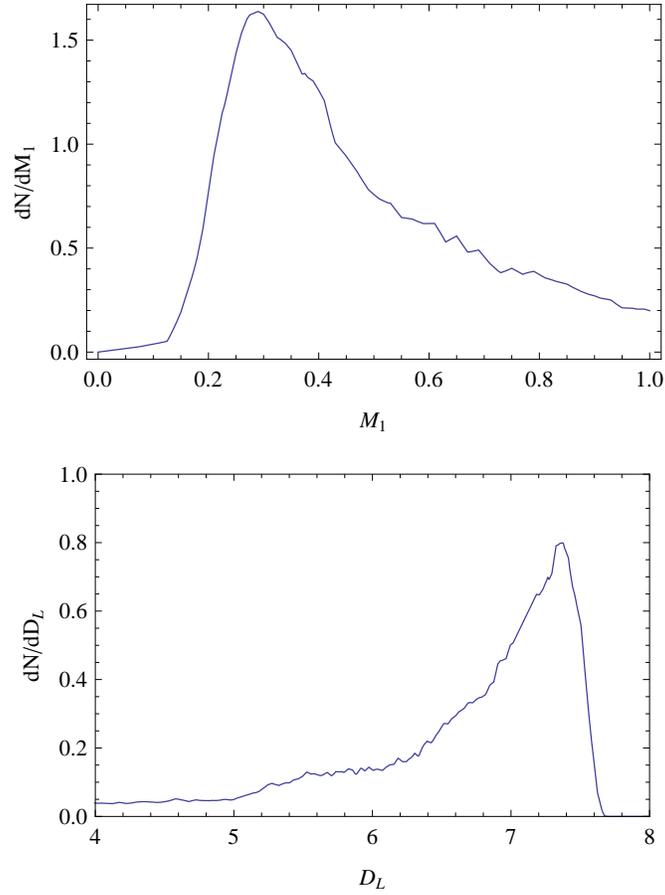}
\caption{Combined probability distributions for the mass of the primary component and for the distance of the lens obtained by weighing the best 8 minima shown in Tables \ref{tab:close} and \ref{tab:wide}.
}
\label{fig:dist}
\end{figure}

Since the microlensing light curve is unable to break the degeneracy
among these eight solutions, the only route we have to a final
statement on the nature of our lens system is to build up a weighted
combination of all probability distributions returned by our Markov
chains. Each probability distribution is weighted by the likelihood
$\exp(-\chi^2/2)$ evaluated on the local maximum and summed to the
others. In the end, we obtain the confidence intervals reported in
Table \ref{tab:phys}. The distributions for the mass of the primary component and the distance to the binary system are shown in Fig. \ref{fig:dist}.

\begin{table}
\centering
\begin{tabular}{cc}
\hline
$\mu$ (mas/yr) & $2.0_{-1.2}^{+6.0}$\\
$M_1/M_\odot$ &  $0.36_{-0.22}^{+1.12}$\\
$M_2/M_\odot$ &  $0.064_{-0.041}^{+0.197}$\\
$D_L$ (kpc) &   $7.18_{-1.68}^{+0.43}$ \\
\hline\end{tabular}
\caption{Physical parameters from the weighted combination of all
minima found.} \label{tab:phys}
\end{table}

The best model indicates a red dwarf as a primary and a brown dwarf as a
secondary. However, due to the concurrence of the unconstrained
minima, the mass ranges of this combined likelihood are much wider
than in the previous tables. In particular, smaller values of $\rho$
correspond to a larger Einstein angle and then a heavier mass and a
smaller distance. In any case, the lens distance distribution still peaks
 as far as 7.375 kpc, which suggests that the lens belongs to the bulge
of our Galaxy. Note that higher masses for the primary at lower
distances would conflict with the constraints from the blending
light. Furthermore, prior expectations favor low-mass lenses, which
are numerically more abundant than higher mass stars.

{\section{Conclusions}
\label{sec:conclude}}

Observations by the Spitzer satellite are rapidly revolutioning the
microlensing field. Traditional ground-based campaigns are
plagued with degeneracies that often remain unsolved with
observations from our planet alone. As a consequence, for some
microlensing events we cannot give a closed scientific
interpretation and we must
complement the models by statistical arguments that combine prior
expectations from our knowledge of the Galaxy. The information on
the presence of planets in microlensing datasets can then be expressed
in terms of probability, which weakens the impact of the potential
discoveries.

With the advent of Spitzer, the situation has radically changed. In
this paper we have seen a clear example in which observations from a
spacecraft far enough from the Earth provide the key to break even
some of the hardest degeneracies in microlensing. Thanks to Spitzer
data, for OGLE-2015-BLG-1212 we have been able to definitely exclude
the presence of a planet, which would have been allowed by ground
data alone. In the path toward the construction of a map of the
planets in our Galaxy, it is crucial that we have the highest
confidence in the interpretation of the microlensing events that we
accept as basic bricks. We have shown that the combination of the ground-based and Spitzer observations is able to establish the nature of
individual microlensing events with unprecedented confidence.

Coming to the details of the modeling of this specific event, we also
note that the measure of such a weak parallax is only possible thanks
to Spitzer and would be impossible using Earth observatories only.
Furthermore, even though the source is not crossing any caustics, it
is very interesting to note that we obtain an upper and a lower limit
for the source size for those models in which the source passes on
the same side of the caustic as seen from Earth and from Spitzer
(models "++" and "--"). This is an unexpected bonus from the presence
of a second probe of the lens plane. Although Spitzer goes further
from the caustic, its light curve still constrains the model in a region of the parameter space in which the ground light curve is better fitted by requiring a minimal size of the source. Summing up, even in the limiting case
of an event far in the bulge with a non-caustic crossing source
trajectory, Spitzer has been able to provide a parallax and an
indication of the source size sufficient to have a clear and complete
idea of the lens system. This is a unique occurrence in the history
of microlensing observations.

With Spitzer observation campaigns, a new era has been opened for the microlensing field. This era is continuing in the next few years with the addition of precious observations by some more satellites already orbiting the Sun or that are being designed at present. In 2016 we will have the Campaign 9 of the K2 mission that will observe the bulge for three months \citep{white}. The separation from Earth is a fraction of AU, thus being comparable to Spitzer's, but this satellite will operate in survey mode, with more than a hundred microlensing events expected. The presence of a sufficiently long baseline for some events will provide an important additional constraint that will be extremely useful in the analysis. If some events will be simultaneously observed from ground, K2 and Spitzer, we will have the incredible possibility to analyze events from three different points of view, which will dramatically reduce the possibilities for degeneracies to survive \citep{SCNScarpetta}. All these observations are a stimulating anticipation of the WFIRST mission\footnote{http://wfirst.gsfc.nasa.gov}, which is specifically designed to perform microlensing searches ten years from now and that will likely yield several hundreds of microlensing planets \citep{Yee14,ZhuGould}. The current design considers a geosynchronous orbit or the Lagrangian point L2. These options would provide a shorter baseline with respect to Spitzer or K2, which would be compensated by a much higher quality of the photometry. The era of microlensing from space has just begun.

\acknowledgments

OGLE Team thanks Profs. M. Kubiak and G. Pietrzy{\'n}ski, former members of the OGLE team, for their contribution to the collection of the OGLE photometric data over the past years. The OGLE project has received funding from the National Science Centre, Poland, grant MAESTRO 2014/14/A/ST9/00121 to AU.

Work by YS and CBH was supported by an appointment to the NASA Postdoctoral Program at the Jet Propulsion Laboratory, administered by Oak Ridge Associated Universities through a contract with NASA.

Work by CH was supported by Creative Research Initiative Program (2009-0081561) of National  Research  Foundation  of  Korea.

JCY, AG, and SCN acknowledge support by JPL grant 1500811. Work by WZ and AG was supported by NSF AST 1516842.

Work by JCY was performed under contract with the California Institute of Technology (Caltech)/Jet Propulsion Laboratory (JPL) funded by NASA through the Sagan Fellowship Program executed by the NASA Exoplanet Science Institute.

TS acknowledges the financial support from the JSPS, JSPS23103002,JSPS24253004 and JSPS26247023. The MOA project is supported by the grant JSPS25103508 and 23340064.

Based on data collected by MiNDSTEp with the Danish 1.54 m telescope at the ESO La Silla observatory.

\end{document}